\documentclass[twocolumn]{aastex631} 

\usepackage{multirow}

\usepackage{graphics,epsf}
\usepackage[utf8]{inputenc}
\usepackage{amsmath}                
\usepackage{amsfonts}               
\usepackage{amssymb}                
\usepackage{epsfig}                 
\usepackage{graphicx}               
\usepackage{float}
\usepackage{color}
\usepackage{multirow}               

\hypersetup{
    colorlinks=true,
    linkcolor=red,   
    urlcolor=cyan}

\usepackage[colorinlistoftodos]{todonotes}

\newcommand{\s}{{~\rm s}}

\newcommand{\g}{{~\rm g}}
\newcommand{\G}{{~\rm G}}
\newcommand{\K}{{~\rm K}}
\newcommand{\erg}{{~\rm erg}}

\begin{document}

\title{The early r-process nucleosynthesis scenarios}

\author[0000-0003-0375-8987]{Noam Soker}
\affiliation{Department of Physics, Technion - Israel Institute of Technology, Haifa, 3200003, Israel; soker@physics.technion.ac.il}
\email{soker@physics.technion.ac.il}

\begin{abstract}
I compare seven actively studied r-process nucleosynthesis scenarios against observed properties of r-process elements in the early Universe, and conclude that the most likely scenario to contribute to the site of elements below the third r-process peak is the magnetorotational r-process scenario, and that of the third peak is the common envelope jets supernova (CEJSN) r-process scenario. The collapsar and CEJSN r-process scenario might also contribute to the lighter r-process elements, and the binary neutron star (NS-NS) merger r-process scenario might contribute to the third r-process peak.  The magnetar, the wind from the newly born NS, and the accretion-induced collapse of a white dwarf r-process scenarios fall short in explaining observations. They might exist, but cannot be major contributors to the r-process in the early Universe.  To constrain r-process scenarios in the early Universe, I require that they explain the large scatter in the r-process abundances of very metal-poor stars, account for the correlation between light r-process nucleosynthesis and iron production, and the lack of correlation between the third peak r-process production and iron production, as inferred from very metal-poor stars. I discuss the diversity of the CEJSN r-process scenario and encourage extending its exploration.  
\end{abstract}


\section{Introduction} 
\label{sec:intro}

The rapid neutron-capture process (r-process), in which intermediate-mass elements capture tens to about 200 neutrons, occurs in dynamically evolving, extremely neutron-rich sites, leading to the nucleosynthesis of the heaviest elements (e.g., \citealt{ThielemannCowan2026}, for a recent review). There are several r-process scenarios: (1) the merger of two neutron stars, the NS-NS merger scenario (e.g., \citealt{Gorielyetal2011, Wanajoetal2014, Beniaminietal2016a, Beniaminietal2016b, BeniaminiPiran2024, Jietal2016, Metzger2017, Banerjeeetal2020,  Dvorkinetal2021, vandevoortetal2022, MaozNakar2025, Qiumuetal2025, Rastinejadetal2025, Zenatietal2026}); (2) the magnetorotational supernova scenario, i.e., core-collapse supernovae (CCSNe) that are powered by a fixed-axis pair of jets (e.g., \citealt{Winteleretal2012, Nishimuraetal2015, Nishimuraetal2017, HaleviMosta2018, Reichertetal2021, Reichertetal2023, Yongetal2021, LiuZetal2025}); (3) the NS winds scenario, i.e., a wind from the newly born NS in a CCSN (e.g., \citealt{Prasannaetal2024}), possibly limited to the first r-process peak, i.e., the weak r-process (e.g., \citealt{WangBurrows2023, WangBurrows2024}), but with the third peak in the presence of magnetars \citep{Prasannaetal2025}; (4) the common envelope jets supernova (CEJSN) r-process scenario (e.g., \citealt{Papishetal2015, GrichenerSoker2019, GrichenerKobayashiSoker2022, GrichenerSoker2022RNAAS, Grichener2023, Grichener2025, Soker2025Rproc}); (5) collapsars, i.e., core-collapse supernovae that forms black holes at the center (e.g.,  \citealt{Siegeletal2019, Siegeletal2022, Braueretal2021, Issaetal2025, Leicesteretal2025, Mumpoweretal2025}), which probably requires significant magnetic fields (\citealt{Justetal2022}); (6) magnetar giant flares (e.g., \citealt{Cehulaetasl2024, Negroetal2025, Pateletal2025a, Pateletal2025b}); (7) accretion induced collapse (AIC) of an ONeMg white dwarf (e.g., \citealt{Batziouetal2025, Cheongetal2025, Combietal2025, Pitiketal2026, Pitiketal2026gama}; in this study, I include the merger of two white dwarfs, i.e., merger induced collapse, under AIC). 

\begin{figure}
\centering
\includegraphics[trim=0.6cm 19.9cm 0.0cm 0.0cm ,clip, scale=0.48]{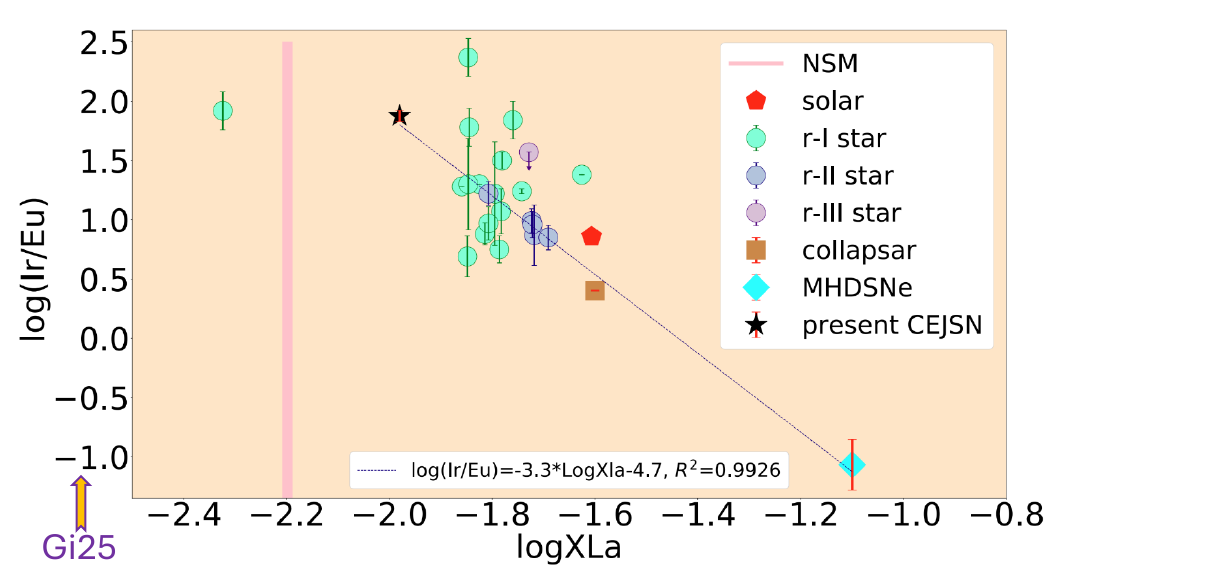}
\caption{A figure adapted from \cite{JinSoker2024} presenting log(Ir/Eu) versus log(XLa), where XLa is the ratio between the lanthanides mass and the total mass of r-process elements. Shown are r-enhanced stars, the solar system, and theoretical calculations of the magnetorotational scenario (marked MHDSNe), the collapsar scenario,  and the CEJSN r-process scenario. Values for r-enhanced stars are from JINAbase (\citealt{Abohalimaetal2018}) with the two constraints of [Fe/H] $<$ -2.5 and [Ba/Eu] $<$ -0.4. For the collapsar model in \cite{Siegeletal2019}, the mean values of the accretion rates $\dot{M}_{1}$ and $\dot{M}_{2}$ are used, which represent the strong r-process components. For the magnetorotational scenario from \cite{Reichertetal2021},  the mean values of the top three strongest r-process traces are presented. The value of $\log{\rm XLa}$ of the NS-NS scenario marked by NSM (pink vertical line) is from \cite{Jietal2019}. I added, in the lower left, the new estimate of the XLa by \cite{Gillandersetal2026} for the kilonova GW170817/AT 2017gfo. The values for the CEJSN scenario are from \cite{JinSoker2024}. The dashed line is a linear fit to the magnetorotational, collapsar, and CEJSN scenarios, with ${R^2}$ = 0.9926.  
}  
\label{fig:xlaIrEu}
\end{figure}

As of today, observations directly and clearly support the NS-NS merger scenario, indicating the presence of very heavy radioactive elements (e.g., \citealt{Chornocketal2017, Kasenetal2017, Pianetal2017, Watsonetal2019, Kasliwaletal2022, Watsonetal2019, Domoto_2022, Levanetal2024}), and possibly in one magnetar giant flare \citep{Pateletal2025a}. However, several papers suggest that the NS-NS merger scenario may not be the sole r-process site (e.g., \citealt{Waxmanetal2018, Coteetal2019, Jietal2019, Kobayashietal2020, Kobayashietal2023, Holmbecketal2024, ChenHYetal2025, Hendersonetal2025ncapture, Saleem2026}; \citealt{ThielemannCowan2026} for a review).
Particularly, studies claimed that the lanthanide fraction (relative to total r-process yield) of kilonova GW170817/AT (SN 2017gfo)  is much below the observed value,  and can be more than an order of magnitude below the observed values: ${\rm XLa} \simeq 10^{-3}$ \citep{Waxmanetal2018}, ${\rm XLa} \approx 6 \times 10^{-3}$ \citep{Jietal2019}, and ${\rm XLa} \approx 2.5 \times 10^{-3}$ \citep{Gillandersetal2026}; most metal-poor stars have $X_{\rm La} \gtrsim 0.015$, as I present in Figure \ref{fig:xlaIrEu} adapted from \cite{JinSoker2024}.   
Many papers discuss the need for two or more r-process sites to explain the Milky Way r-process abundance (e.g., \citealt{Wehmeyeretal2015, GrichenerSoker2019paradigm, HaynesKobayashi2019, Moleroetal2021, Tsujimoto2021, Farouqietal2022, Naiduetal2022, Yamazakietal2022, Stormetal2025}). \cite{Rastinejadetal2025} who study the NS-NS merger scenario, find that there is room for more r-process sites in addition to the NS-NS merger. 

Some recent studies of r-process abundance in metal-poor stars claim that at least two r-process sites should have operated in the young Galaxy and Universe (e.g., \citealt{Farouqietal2025, Kuskeetal2025, Sarafetal2025Diff, Zenatietal2026}). Other recent studies examine the scatter in r-process abundance among different metal-poor stars (e.g., \citealt{Bandyopadhyayetal2024, Hansenetal2024, XylakisDornbuschetal2024, Griffithetal2026, Sarafetal2026}). I summarize the relevant new findings of these studies (Section \ref{sec:Abundance}) and compare the different theoretical r-process scenarios with these new observations (Section \ref{sec:Scenarios}). 

The r-process scenarios are a hot topic, with numerous recent papers addressing this topic (e.g., \citealt{Hayesetal2023, Placcoetal2023,  Bishopetal2025, Hansenetal2025, HiraiYetal2025, Kobayashietal2025, MaozNakar2025, Moleroetal2025, Pallaetal2025,  Stormetal2025, TianMetal2025, XieXetal2024, XieXetal2025, Anoardoetal2026, Santarellietal2026}). 
The goal of this paper is to examine actively studied r-process sites in light of recent studies of r-process nucleosynthesis in metal-poor stars, particularly in the early Universe. Namely, I address one question: Which r-process scenarios can contribute to the r-process in extremely metal-poor stars? Although this study presents no new calculations, it is the only study to systematically compare all actively studied r-process scenarios, omitting none; this allows me to conclude which scenarios are likely to contribute to r-processes in the early Universe.  
Therefore, it offers a timely summary of new results from 2025 and 2026, and a valuable comparison of theoretical scenarios with observed r-process element properties in the early Universe. The conclusions, summarized in Section \ref{sec:Summary}, have significant implications for future theoretical studies of r-process scenarios.

\section{R-process in metal-poor stars} 
\label{sec:Abundance}

To achieve my goal of listing the possible scenarios that most strongly contribute to r-process nucleosynthesis in the early Universe, I focus on two observational results from the literature, particularly those published in 2025-2026. 

\subsection{Large abundance scatter} 
\label{subsec:scatter}

\begin{figure}
\begin{center}
\includegraphics[trim=0.4cm 15.1cm 0.0cm 0.0cm ,clip, scale=0.42]{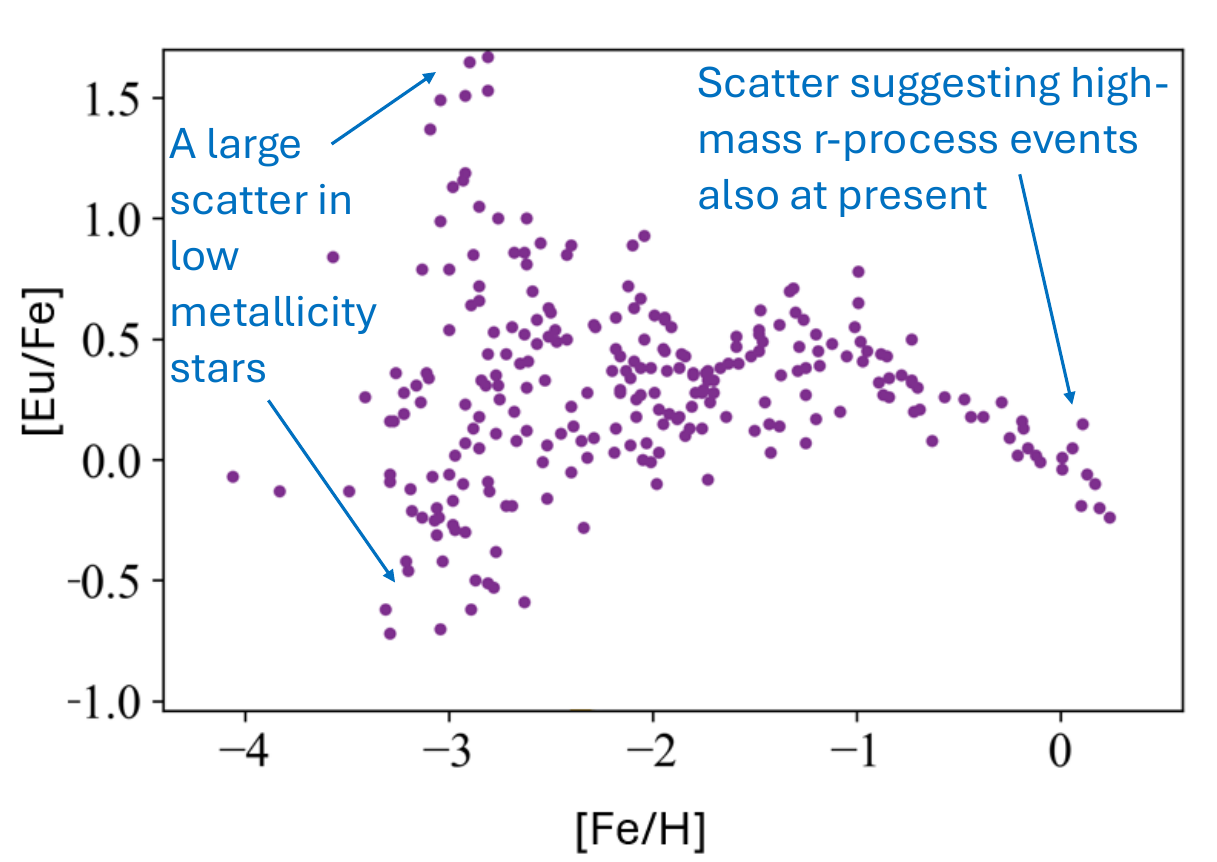}
\caption{A figure demonstrating the large scatter in r-process nucleosynthesis at early times with the ratio [Eu/Fe]. The scatter at late times suggests that r-process sites that yield a large mass of r-process elements also dominate at later times (i.e., present). Observational data taken from \cite{Cayreletal2004}, \cite{Hondaetal2004}, \cite{Hansenetal2012}, \cite{Hansenetal2014}, \cite{Roedereretal2014} and \cite{Zhaoetal2016} (figure based on the one in \citealt{Kobayashietal2020} and \citealt{GrichenerKobayashiSoker2022}).      
}
\label{fig:RPscatter}
\end{center}
\end{figure}

From very early times of having [Fe/H]$\lesssim -3$, the abundance of r-process elements, like Eu, shows very large scatter, as I reproduce in Figure \ref{fig:RPscatter} based on \cite{Kobayashietal2020}. This shows that there is no averaging over many r-process nucleosynthesis events at those early times. Therefore, each event must produce a relatively large mass of r-process elements. From [Fe/H]$\simeq -3$, the scatter decreases with increasing metallicity (e.g., \citealt{Ouetal2025}), indicating the averaging over more and more r-process events; above ${\rm [Fe/H]} \gtrsim -1.5$, iron production in type Ia supernovae also reduces the scatter. \cite{Hendersonetal2025} studied 89 stars in the globular cluster M15 and found high Ba, La, and Eu dispersions in the first generation of stars. They conclude, under the assumption that the r-process events that caused the abundance dispersions were born with the first population of stars in M15, that the r-process site must have a short delay time.

These and other studies (Section \ref{sec:intro}) show that all stars contain iron, which at these early times is produced by CCSNe. Iron had time to well mix with the ISM that forms these old stars. The presence of iron and the large scatter in r-process elements indicate that r-process sites are very rare relative to CCSNe. On the other hand, the presence of r-process elements indicates that the delay between CCSNe and r-process events is very short.

\cite{Raccaetal2025} analyzed the abundance of 10 metal-poor stars and found small abundance dispersions. They concluded that this suggests a high degree of uniformity in r-process yields across diverse astrophysical environments. These stars also show large scatter in [Eu/Fe]. 

\cite{Kobayashietal2020} argue that $\simeq 3\%$ of $25 - 50 M_\odot$ that are magnetorotational supernovae that produce r-process elements can explain the r-process they deduce for the Galaxy. 

\cite{Jietal2023} argued that a single r-process event enriched the ultrafaint dwarf galaxy Reticulum II, and that this single event produced an r-process element mass of $M_{\rm r} \simeq 0.06 - 0.6 M_\odot$. 
More generally, they argued that the early r-process site(s) are prompt and have high yield. 

The well-established conclusions from the above-cited papers and many more are as follows:  (1) R-process events in the early Universe were very rare; hence, (2) they must have had a massive r-process yield per system, $M_{\rm rp} > 0.01 M_\odot$, and possibly a few times larger. (3) The r-process events must have a short delay from the first CCSNe.  

\subsection{Two early r-process sites} 
\label{subsec:TwoSites}

\cite{Farouqietal2022} argue that at least three r-process nucleosynthesis sites were active in the early Galaxy: two produced Fe and the majority of the lighter r-process elements. In contrast, the third produced strong r-processed (third peak) elements. They attributed the first two to two distinct types of CCSNe, but not to standard CCSNe, and the third to NS-NS mergers, some of which rapidly produce a black hole, and some do not. 

\cite{Farouqietal2025} find that the Eu and Fe are co-produced at the early Universe, while the third r-process peak elements are not co-produced with Fe; the third r-process peak and Eu production are decoupled (e.g., \citealt{AlencastroPulsetal2025}).
\cite{Sarafetal2025Diff} claimed the necessity of at least two r-process sites. They found that the NS-NS merger scenario cannot account for the abundance of stars with [Sr/Eu]$\gtrsim -0.7$, which also tend to have a low value of [Eu/Fe]. However, they also find a continuous distribution of metal-poor stars in the plane of [Sr/Eu] versus [Eu/Fe]. Their finding is compatible with the claim of \cite{Farouqietal2025} that Eu and Fe are coproduced. 

I adopt these recent claims for two early r-process site types: one for light r-process elements near europium and lighter, which is correlated with iron production (I term it ``Eu''+Fe), and the other for the third-peak r-process nucleosynthesis. 

\section{Scenarios comparison} 
\label{sec:Scenarios}

In Table \ref{Tab:Table1}, I compare the seven scenarios listed in Section \ref{sec:intro} with respect to their ability to explain the observations related to the early r-process in the Universe described in Section \ref{sec:Abundance}. I consider two types of r-process sites (Section \ref{subsec:TwoSites}): one that produces the light elements in the early universe, with some correlation with iron production (I term it ``Eu''+Fe), and the other that forms the third r-process peak. I also require that the scenarios operate with a short delay after the first CCSNe and account for the large scatter in r-process elements (Section \ref{subsec:scatter}). A `$(-)$' symbol indicates that the scenario is not a main contributor to the nucleosynthesis site in the early universe. It does not imply that it does not occur. I cannot rule out that it occurs, but only that it is not the main contributor. Three of the scenarios predict that the r-process event occurs long after the CCSN; hence, in these scenarios, r-process elements are generally not correlated with iron production in the early Universe (before type Ia supernovae occur). These are indicated by `No' in the third column. The other four occur alongside the CCSN explosion (or starting within seconds after the explosion).     
\begin{table*}
\begin{center}
  \caption{Comparing r-process scenarios in the early Universe}
  \begin{tabular}{|p{2.5cm} | p{2.5cm}| p{0.5cm}| p{5.0cm}| p{5.0cm}|}
\hline   
 \textbf{Scenario} & \textbf{RP yield} & \textbf{Fe} & \textbf{R-process: ``Eu''+Fe } & \textbf{R-process: third peak } \\ 
\hline  
\hline 
 NS-NS merger & $\approx 0.03 M_\odot$ \textcolor{blue}{$^{\rm [Ra18]}$} & No & \textcolor{red}{($-$) No Fe expected$^{[\&]}$} & ($=$) Ultrafast mergers\textcolor{blue}{$^{\rm [BP24]}$}; unclear if sufficient number of systems   \\ 
 \hline  
 Magnetorotational & $\simeq 0.003-0.03 M_\odot$ \textcolor{blue}{$^{\rm [Mo18]}$} &  Yes & \textcolor{green! 65! black}{($+$) Possible} & \textcolor{red}{($-$) Correlated with Fe} \\ 
\hline  
 NS or magnetar wind & $\approx 10^{-4} M_\odot$ \textcolor{blue}{$^{\rm [Pr25]}$}  & Yes & \textcolor{red}{($-$) Too little mass to explain scatter} & \textcolor{red}{($-$) Too little mass to explain scatter}; \textcolor{red}{($-$) Correlated with Fe}\\ 
  \hline  
 CEJSN r-process& $\simeq 0.01-0.05 M_\odot$ \textcolor{blue}{$^{\rm [G9S5]}$} & No & $(=)$ In rare cases of a very massive oxygen core\textcolor{blue}{$^{\rm [Section~ \ref{sec:Summary}]}$} & \textcolor{green! 65! black}{($+$) Possible: like a large ratio of Ir/Eu predicted}\textcolor{blue}{$^{\rm [JS24]}$} \\ 
  \hline  
 Collapsar & $\simeq 0.08- 0.3 M_\odot$ \textcolor{blue}{$^{\rm [Si19]}$} & Yes & ($=$) Only very strict conditions might yield sufficient mass\textcolor{blue}{$^{\rm [Is25]}$} & \textcolor{red}{($-$) Correlated with Fe} \\ 
  \hline  
 Magnetar flares & $\simeq 10^{-5} - 10^{-3} M_\odot$ \textcolor{blue}{$^{\rm [Pa25]}$} & Yes & \textcolor{red}{($-$) Too little mass to explain scatter} & \textcolor{red}{($-$) Too little mass to explain scatter}  \\ 
  \hline  
 AIC  & $\approx  10^{-2} M_\odot$ \textcolor{blue}{$^{\rm [Co25]}$}; $\approx  0.1 M_\odot$ \textcolor{blue}{$^{\rm [Pi26]}$} & No &  \textcolor{red}{($-$) Too long delay} &  \textcolor{red}{($-$) Too long delay} \\ 
\hline  
     \end{tabular}
  \label{Tab:Table1}\\
\end{center}
\begin{flushleft}
\small 
Notes: Comparing the different r-process scenarios during the early Universe. Additional references with further information on the properties of the different scenarios are provided in Section \ref{sec:intro}. A `$(-)$' symbol indicates that the scenario is not the main contributor to the nucleosynthesis site in the early universe. It does not imply that it does not occur. I cannot rule out that it occurs, but only that it does not contribute substantially. 
\newline 
Columns: (1) The r-process scenario; (2) The mass of the r-process elements that one system produces; (3) The correlation with iron: A `Yes' indicates that the scenario occurs during or immediately after a CCSN where iron is produced. (4) My estimate of whether a scenario can be a major contributor to the nucleosynthesis of light r-process elements (below the third peak), such as Eu, that are observed to correlate with iron in the early Universe (Section \ref{subsec:TwoSites}); (5) My estimate of whether a scenario can be a major contributor to the third peak in the early Universe. 
\newline
Comment: [\&] Recently, \cite{JacobiMetal2026} find that some NS-NS merger can yield $\simeq 4 \times 10^{-4} - 8 \times 10^{-4} M_\odot$ of $^{56}$Ni. This is about two orders of magnitude less than the yield of a single CEJSN. 
\newline
Abbreviations: AIC: accretion-induced collapse; CEJSN: common envelope jets supernova; NS: neutron star; RP: r-process. 
\newline 
References:  
BP24: \cite{BeniaminiPiran2024} and \cite{MaozNakar2025}; 
Co25: \cite{Combietal2025}; 
G9S5: \cite{GrichenerSoker2019} and \cite{Soker2025Rproc}; 
Is25: \cite{Issaetal2025}; 
JS24: \cite{JinSoker2024};
Mo18: \cite{Mostaetal2018};
Pa25: \cite{Pateletal2025a}; 
Pi26: \cite{Pitiketal2026,Pitiketal2026gama};
Pr25: \cite{Prasannaetal2025}; 
Ra18: \cite{Radiceetal2018}; 
Si19: \cite{Siegeletal2019}; 
\end{flushleft}
\end{table*}

The NS-NS merger scenario can have a long delay after two CCSNe. \cite{Kobayashietal2023}, for example, argued that the delay time of NS-NS merger (as well as black hole-NS merger) is too long to explain the early Galactic r-process. On the other hand, \cite{BeniaminiPiran2024} claimed that the natal kick of the second NS can lead to ultrafast merger, which can explain early r-process in the Universe (for early NS-NS mergers see also \citealt{MaozNakar2025}). This scenario definitely contributes at late times, and might contribute some fraction of the third peak at early times. However, I do not currently consider it a major contributor to the third peak at early times. But this is an open question. Another recent problem is the rate of NS-NS merger, which is below earlier estimates from gravitational wave sources, and might be below the one required to explain all r-process isotopes \citep{Fishbachetal2026}.

\cite{Okadaetal2026} studied an extremely metal-poor star, [Fe/H]$\simeq -4$, and concluded that the sharp decline in abundances beyond Zr disfavors the NS-NS merger scenario and favors the magnetorotational supernova scenario. They did not study the third r-process peak. I consider the magnetorotational supernova scenario as a possible major contributor to the nucleosynthesis of the light r-process elements in the early Universe, alongside the collapsar scenario. The major difference between these two scenarios is that in the first one, accretion is onto an NS remnant, while in the second, the accretion is onto a black hole. 

\cite{Prasannaetal2025} studied the r-process in the wind from a newly born magnetar. They argue that it can be a significant site for the r-process. They estimate the r-process yield per event to be $M_{\rm r,ev} \approx 10^{-4} M_\odot$. This scenario cannot account for the large scatter at early times, as it cannot be a rare event relative to CCSNe to explain the early r-process elements. The same holds for the r-process from a quark nova, an event where an NS experiences a phase transition to a quark star and ejects neutron-rich mass, because the r-process yield is only $M_{\rm r,ev} \approx  10^{-5} - 10^{-4} M_\odot$ (e.g., \citealt{Ouyedetal2009, Ouyed2022}). 

Because studies much too often ignore the CEJSN r-process scenario, and because I think it is the major contributor to the third peak at early times, I elaborate a little on this scenario here, and in Section \ref{sec:Summary}, where I speculate on the possibility that this scenario also contributes to the ``Eu''+Fe site. 
The CEJSN r-process scenario (e.g., \citealt{GrichenerSoker2019, Grichener2025}) can account for the europium evolution in the Galaxy \citep{GrichenerKobayashiSoker2022}, the nucleosynthesis of the three r-process peaks \citep{JinSoker2024}, and other observed r-process properties (e.g., \citealt{GrichenerSoker2019paradigm, Grichener2023}).  
In the CEJSN r-process scenario, an NS enters the core of a massive star, accretes mass at a high rate, and launches jets. The jets expel some core material; the rest accretes onto the NS, forming a dense accretion disk that produces neutron-rich material that drives the r-process in the disk and at the base of the jets. Mixing of disk material with the neutron-rich crust of the NS increases the neutron fraction (lowering $Y_{\rm e}$; \citealt{Soker2025Rproc}), and can increase the total r-process elements yield beyond earlier estimates \citep{GrichenerSoker2019, GrichenerSoker2019paradigm}. The jets carry the r-process elements away. 
 In a very recent study, \cite{ShahSetal2026} find that in metal-poor stars, there is a relatively small scatter in the ratio of Th to the lanthanides Eu and Dy. They conclude that existing models struggle to explain both prompt r-process and a robust Th/Eu ratio. I speculate here that the mixing layer of the accretion disk and the NS in the CEJSN r-process scenario can account for both. The CEJSN r-process is a prompt site, and the mixing layer between the accretion disk and the NS is expected to be similar across events and to have a low $Y_{\rm e}  < 0.2$. Their finding might suggest that, in the CEJSN r-process scenario, the jets originating from the mixing layer carry most of the r-process elements.   

 I consider the CEJSN r-process scenario the primary contributor to the third r-process peak for the following reasons.  
(1) The ratio of (Ir/Eu), i.e., third peak to lighter r-process, of the CEJSN r-process site as calculated by \cite{JinSoker2024}, is more than an order of magnitude above that of the collapsar (taken from \citealt{Siegeletal2019}) and about three orders of magnitude above that of the magnetorotational scenario (taken from \citealt{Reichertetal2021}); see Figure \ref{fig:xlaIrEu}. (2) The CEJSN events are much less associated with the location of CCSNe, as there is a delay from the CCSN explosion that forms the NS to the merger event. The events associated with CCSNe (i.e., magnetorotational, NS winds, collapsars, and young magnetar flares) have shorter delays, are likely to occur in stellar clusters where there are many CCSNe, and might produce large amounts of Fe because they are CCSNe. (3) Despite not being at a CCSN location, the delay of the CEJSN r-process is very short, as it occurs before the second massive star finishes its regular evolution.    

I note that in a recent paper \cite{HallSmithetal2026} propose that when an NS accretes mass in a CEJSN impostor (impostor implies that the NS is in the envelope rather than in the core), the lower accretion rate can form proton-rich ejecta, which leads to the rapid proton (rp) process. This CEJSN impostor rp-process scenario needs further study and comparison with the CEJSN r-process scenario. 

\cite{Siegeletal2019} argued that the collapsar r-process scenario, in which a newly born black hole accretes from a collapsing core, can produce large amounts of r-process elements. However, it has some problems (e.g., \citealt{BartosMarka2019}). For example, it requires very strong magnetic fields and very high accretion rates onto the black hole ($>10 M_\odot \s^{-1}$;  \citealt{Issaetal2025}). This seems too high for the outer core, since the inner core collapsed to form the black hole and is accreting mass from the outer core's zones. I consider this scenario as a less likely contributor. This requires further studies \citep{Issaetal2025}. 

Magnetars produce $10^{-5} - 0.001 M_\odot$ of r-process elements per magnetar \citep{Pateletal2025a, Pateletal2025b}, which is too little mass and cannot explain the scatter. Namely, the small mass implies that many events are already required at an early stage, hence not a large random distribution of r-process elements in the early Universe.   

The AIC r-process scenario appears only later in Galactic and Universe history, and cannot be a major contributor in the early Universe.

\section{Discussion and Summary} 
\label{sec:Summary}

I considered seven actively studied r-process scenarios (first column of Table \ref{Tab:Table1}; \citealt{ThielemannCowan2026} present a review of different aspects of the different r-process sites). I focused on two of their properties: the yield of r-process elements (second column of Table \ref{Tab:Table1}) and whether nucleosynthesis occurs alongside iron nucleosynthesis, i.e., alongside a CCSN (third column of Table \ref{Tab:Table1}). I estimated their ability to account for two types of r-process sites (Section \ref{subsec:TwoSites}): One type synthesis mainly elements below the third peak with some correlation with iron (marked ``Eu''+Fe in the fourth column of Table \ref{Tab:Table1}), and the second that synthesizes mainly the third peak of the r-process. 
I concluded that the magnetorotational r-process scenario is most likely to be the major contributor to the ``Eu''+Fe sites, but that the collapsar r-process scenario might also contribute some (fourth column of Table \ref{Tab:Table1}). I also concluded that the CEJSN r-process scenario is the primary contributor to the third peak, but that the NS-NS merger scenario may also contribute (see the fifth column of Table \ref{Tab:Table1}).
There are two important differences between the collapsar scenario, in which accretion is onto a black hole, and the magnetorotational and CEJSN r-process scenarios, in which accretion is onto an NS \citep{Soker2025Rproc}.  (1) The solid surface implies a boundary layer within the inner accretion disk that may provide conditions conducive to the formation of a neutron-rich gas. (2) In addition, the boundary layer can mix neutron-rich material from the NS crust.  

In the CEJSN r-process scenario, the NS accretes at a very high rate as it enters the core of a massive star (e.g., \citealt{GrichenerSoker2019}). Following mass accretion and r-process nucleosynthesis driven by the jets, the NS will spin at a very high rate. This might drive strong winds and magnetar flares, which in turn could drive further r-process nucleosynthesis, but at a much smaller yield (see Table \ref{Tab:Table1}). Alternatively, the NS might collapse to a black hole during the accretion process, such that the last phase of the high-accretion rate occurs onto a black hole. 
To explain actinide boost stars, \cite{Farouqietal2022} proposed that the NS-NS merger site involves two sub-channels: one that immediately forms a black hole with a torus around it, similar to a black hole-NS merger event, and one where most accretion after merger is onto a massive NS. The same splitting might occur in the CEJSN r-process scenario: The NS that enters the massive core might accrete via an accretion disk and launch jets, and then collapse to a black hole, so that accretion proceeds onto a black hole. The continuation accretion onto the black hole might increase the estimated r-process yield of the CEJSN r-process scenario and bring it to $\gtrsim 0.1 M_\odot$.  

Here, I speculate that the CEJSN r-process might also contribute to the "Eu"+Fe site. In the basic CEJSN r-process scenario, the NS accretes mass and launches jets that expand almost freely. The jets do not shock the core material to the extent that iron production occurs, as in CCSNe. This is because, as the NS spirals into the massive core, the jets and gravitational force of the NS destroy the core to form an accretion disk around the NS. However, there are rare cases, which might have been more common in low-metallicity stars in the early Universe when stars were more massive, that the oxygen core is very massive, $M_{\rm O-core} \gtrsim 15 M_\odot$. Such stripped-envelope cores are descendants of zero-age main-sequence stars of $M_{\rm ZAMS} \gtrsim 40 M_\odot$. In the binary system, the primary's mass was lower, and the secondary's mass was lower as well; however, mass was transferred from the primary to the secondary, which increased the secondary's mass to the required value to form a massive core. The binding energy of such a core is $E_{\rm core,b} \gtrsim 2 \times 10^{51} \erg$. In those extreme cases, the NS and its jets do not immediately destroy the core, and the NS spirals deep into the core. Due to the large angular momentum, the flow now resembles that in the magnetorotational r-process scenario, with one major difference: the NS in the CEJSN r-process scenario is cold. The newly born NS in the magnetorotational r-process scenario emits ${\rm several} \times 10^{53} \erg$ in neutrino-anti-neutrino pairs, which act against high neutron fraction due to the reaction $n+\nu \rightarrow p + e^{-}$. In case the neutron accretes sufficient mass, it collapses to a black hole, after which the flow resembles that of the collapsar r-process scenario. The jets now shock the oxygen-rich core material and may drive nucleosynthesis of iron. Therefore, the CEJSN r-process scenario might also contribute to the ``Eu''+Fe site. This requires detailed simulations, which I highly encourage. 

\section*{Acknowledgments} 
I thank Aldana Grichener for enlightening conversations and suggestions. I thank the Charles Wolfson Academic Chair at the Technion for the support.

\bibliography{reference}{}
\bibliographystyle{aasjournal}

\end{document}